\documentclass[journal,twocolumn]{IEEEtran}
\usepackage{amsmath,amsfonts,amssymb}
\usepackage{comment}
\usepackage{algorithmic}
\usepackage{algorithm}
\usepackage{array}
\usepackage{textcomp}
\usepackage{graphicx}
\usepackage{subfigure}
\usepackage{cite}
\newtheorem{proposition}{Proposition}
\begin{document}

\title{Model-Order-Adaptive Channel Estimation for AFDM Systems with Fractional Delay and Doppler}
\author{Ruizhe Wang, Cunhua Pan, Hong Ren, Shaohua Li, and Jiangzhou Wang, \IEEEmembership{Fellow, IEEE}
		\thanks{
		The authors are with National Mobile Communications Research Laboratory, Southeast University, Nanjing, China. (e-mail: {rzw, cpan, hren,  230268176, j.z.wang}@seu.edu.cn).

		\emph{Corresponding author: Cunhua Pan and Hong Ren}}
	}

\maketitle

\begin{abstract}
Affine frequency division multiplexing (AFDM) has emerged as a promising waveform for high-mobility communications owing to its ability to exploit multipath diversity. However, channel state information (CSI) acquisition remains challenging in channels with fractional delay and Doppler and an unknown number of propagation paths. In this paper, we investigate the channel estimation for AFDM systems. 
We first analyze the AFDM response in the presence of fractional-delay-induced frequency wrapping. The analysis reveals that fractional delay may displace the dominant extremum and generate informative secondary extrema. 
Then, we develop a model-order-adaptive channel estimator within an improved space-alternating generalized expectation-maximization (SAGE) framework. A dual-pilot reference symbol is employed for path and delay initialization, while pilot observations across multiple AFDM symbols provide temporal information for Doppler estimation. New paths are identified through statistically controlled residual tests, whereas unsupported or redundant paths are removed by conditional support pruning. The delay, Doppler frequency, and complex gain of each retained path are subsequently estimated through a coarse-to-fine procedure and refined using the exact AFDM likelihood. Simulation results demonstrate that the proposed method achieves lower normalized mean squared error (NMSE) and bit error rate (BER) than representative SAGE, sparse-recovery, and Bayesian benchmarks. It also provides reliable path detection and model-order estimation and maintains robust performance at terminal velocities of up to 600~km/h.
\end{abstract}


\section{Introduction}
With the rapid evolution of next-generation wireless networks, ubiquitous connectivity has emerged as a fundamental requirement to deliver seamless and reliable services spanning terrestrial, aerial, maritime, and satellite domains \cite{ITU-R-M2160}. This demand is further fueled by emerging high-mobility use cases, such as high-speed railways, vehicle-to-everything (V2X) communications, unmanned aerial vehicles (UAVs), and low-Earth-orbit (LEO) satellite systems. In these scenarios, terminals need to sustain uninterrupted access to broadband, low-latency services. Nevertheless, fast relative motion between transceivers and scattering objects introduces significant Doppler shifts and rapidly time-varying multipath channels, which leads to severe time-frequency dispersion. Such channel features bring considerable challenges to traditional wireless waveforms and channel estimation schemes, including orthogonal frequency division multiplexing (OFDM) \cite{ofdmdopplerspread}. Hence, enabling reliable transmission over doubly dispersive channels is critical to achieving global ubiquitous connectivity for future wireless networks.

To improve the reliability of wireless transmission over doubly dispersive channels, several waveforms have been developed for high-mobility communications. Orthogonal time frequency space (OTFS) modulation maps information symbols onto the delay-Doppler (DD) domain and spreads each symbol over the entire time-frequency plane, thereby exploiting the diversity provided by time-varying multipath channels \cite{Hadani2017OTFS}. More recently, orthogonal delay-Doppler division multiplexing (ODDM) has been proposed to construct a realizable waveform that preserves orthogonality with respect to the fine delay and Doppler resolutions of the transmitted frame \cite{Lin2022ODDM}. Although OTFS and ODDM provide a DD domain representation of the channel, their implementations generally involve two-dimensional signal processing, while their performance also depends on appropriate pulse design and DD domain channel estimation and detection.

Against this background, affine frequency division multiplexing (AFDM) has emerged as a promising chirp-based waveform for high-mobility communications \cite{bemani2023afdm,AFDM}. The information symbols are modulated through the discrete affine Fourier transform (DAFT). The chirp parameters can be adapted to the delay and Doppler spreads of the channel. 
Moreover, AFDM characterizes the doubly dispersive channel within a one-dimensional transform. 
With appropriately selected chirp parameters, the DAFT-domain channel provides a full DD representation and allows different propagation paths to be distinguished. Consequently, AFDM can theoretically achieve the optimal diversity order in doubly dispersive channels.

The performance advantages of AFDM, however, rely on accurate channel state information (CSI) at the receiver. In practical high-mobility channels, the path delays and Doppler shifts are generally not aligned with the sampling grid. The resulting fractional delay and Doppler components spread the energy of each path over multiple DAFT-domain samples and introduce interference among adjacent path responses. Accurate estimation of the continuous delay, Doppler, and complex gain of each path is therefore essential for exploiting the delay-Doppler representation and diversity offered by AFDM.

Pilot-aided channel estimation and sensing methods have been developed for AFDM systems, including embedded-pilot channel estimation, single- and multiple-pilot-aided estimation, guard-interval-free iterative estimation, and time- or DAFT-domain target sensing \cite{bemani2023afdm,Yin2022AFDMCE,Zhou2024GIFree,Ni2022AFDMISAC,Bemani2024AFDMISAC}. Nevertheless, these methods mainly rely on discrete-index detection or grid-based searches, and their capability to jointly resolve continuous fractional delays and Doppler frequencies is therefore limited by the adopted grid resolution.
The work in \cite{Cao2024AFDMOMP} investigated AFDM transmission over multi-scale multi-lag underwater acoustic channels and proposed an iterative multi-index estimator for low-to-moderate Doppler time scaling alongside an orthogonal matching pursuit (OMP)-based estimator for severe Doppler time scaling. 
To further exploit the inherent sparsity of AFDM channels, SBL-based methods have been introduced for sparse  parameter estimation and channel recovery \cite{Yang2024AFDMSBL,Luo2025AFDMISACSBL,Tang2025AFDMLaplaceSBL,Lin2025AFDMGAMPSBL,Li2026AFDMIoVSBL,Xu2026AFDMPNSBL,Luo2026AFDMDGESBL}.
Nevertheless, these methods still formulate the estimation problem using a fixed or iteratively adjusted DD domain dictionary. 

Specifically, fractional delay coupled with wrapping introduces segment-dependent phase terms. Hence, the DAFT-domain response cannot be reduced to a simple superposition of conventional Dirichlet kernels, but corresponds to an intricate sum of piecewise geometric series.
The work in \cite{chen2026} investigated joint fractional delay and Doppler estimation for AFDM-enabled vehicular LEO-ICAN systems under spectrum wrapping and developed a low-complexity estimator by exploiting the spectrum-wrapping-induced envelope structure.
In \cite{Li2026MF}, the authors proposed a matched-filtering estimators with generalized Fibonacci search. However, path extraction is terminated using a preset maximum iteration count and an empirical residual-change threshold, rather than an explicit adaptive model-order criterion.

Model-order selection has been widely studied in spectral estimation and multipath channel estimation. Information-criterion and residual-testing methods determine the model order by penalizing unnecessary components or accepting new components according to statistically defined thresholds \cite{Xiong2017BIC,Mamandipoor2016NOMP,Pinto2026RobustSAGE}. Alternatively, sparse Bayesian methods infer the active components through sparsity-promoting priors, posterior updates, and adaptive pruning \cite{Shutin2011VBSAGE,Shutin2015IARD,Wang2026VBSAGE}.
Nevertheless, information-criterion methods may require parameter estimation for multiple candidate orders, while residual tests and Bayesian pruning depend on the noise model, prior assumptions, or prescribed thresholds. Moreover, these methods are mainly developed for conventional sinusoidal or OFDM response models and do not account for the coupled fractional delay and Doppler parameters and modulo-induced frequency wrapping of AFDM signals.

In AFDM systems, parameter estimation and model-order determination are closely coupled. An inaccurate parameter estimate leaves structured residual energy and may split one physical path into multiple components, whereas premature stopping or pruning may remove weak paths. Motivated by this issue, we incorporate statistically controlled path initialization and conditional pruning into an improved SAGE framework, where the unknown model order and continuous path parameters are jointly determined using the exact AFDM response, dual-pilot observations, and temporal information across multiple symbols. 
The main contributions of this paper are summarized as follows:
\begin{itemize}
\item
We analyze the exact input--output relationship of AFDM systems in the presence of fractional delay and Doppler. Specifically, we investigate the modulo-induced frequency wrapping and analyze the extrema of the resulting AFDM response. Our analysis shows that fractional delay can displace the dominant extremum and produce informative secondary extrema, thereby providing a basis for subsequent parameter initialization and refinement.
\item
We develop a channel estimation scheme that employs dual pilots in the reference symbol for parameter initialization and exploits temporal pilot observations across multiple AFDM symbols for subsequent refinement. 
A coarse-to-fine parameter estimation procedure is proposed, in which the dominant and secondary extrema provide the initial delay hypotheses, while the fractional delay and Doppler are subsequently refined using the exact AFDM likelihood. The temporal phase variations across symbols are further utilized to improve Doppler estimation, followed by least-squares estimation of the complex path gains.
\item
To address the unknown number of propagation paths, we incorporate statistically controlled path initialization and conditional support pruning into an improved space-alternating generalized expectation-maximization (SAGE) framework. 
New paths are successively extracted from the residual observations, whereas unsupported or redundant paths are removed according to their residual contribution.
Therefore, the model order and the continuous path parameters can be jointly determined without evaluating multiple prescribed model orders. 
The dominant computational complexity of the proposed algorithm is analyzed.
\item
Simulation results demonstrate that the proposed method achieves superior communication performance compared with benchmarks. In particular, it provides better NMSE performance and BER performance, reliable path detection and model-order estimation, and robust performance when the terminal velocity increases from 300~km/h to 600~km/h. 
Moreover, its path-detection F1 score improves steadily with the signal-to-noise-ratio (SNR), while the estimated model order gradually approaches the true average number of propagation paths. 
These results verify that the proposed path initialization and pruning strategy effectively balances missed detections and false alarms.
\end{itemize}

The remainder of this paper is organized as follows. Section~\ref{sec:system_model} presents the AFDM system model and derives the exact input--output relationship in the presence of fractional delay and Doppler. Section~\ref{sec:atom_extrema} analyzes the extrema and identifiability of the wrapped AFDM response. Section~\ref{sec:proposed_ce} develops the proposed adaptive model-order SAGE channel estimation algorithm. Its monotonicity and computational complexity are discussed in Sections~\ref{sec:monotonicity} and~\ref{subsec:proposed_complexity}, respectively. Section~\ref{sec:simulation_results} presents the simulation results and evaluates the proposed method against the benchmark schemes. Finally, Section~\ref{sec:conclusion} concludes this paper.

\section{System Model}
\label{sec:system_model}

\subsection{Single-Antenna AFDM Transmission}

Consider a downlink AFDM scenario in which the base station (BS) and the mobile station (MS) are each equipped with a single antenna.\footnote{The proposed estimator can be extended to multi-antenna AFDM systems by incorporating the spatial response of each propagation path; the subsequent delay--Doppler estimation procedure remains unchanged.} The system employs $M$ subcarriers in total. The sampling interval is defined as $T_s=\frac{1}{M\Delta f}$, where $\Delta f$ stands for the subcarrier spacing. The chirp-periodic prefix (CPP) occupies a duration of $T_{\mathrm{CPP}}=M_{\mathrm{CPP}}T_s$. We use $T_{\mathrm{sym}}$ and $T$ to represent the symbol duration with and without the CPP, respectively. It follows that the symbol duration excluding the CPP satisfies $T=\frac{1}{\Delta f}$ and $T=\frac{M}{M+M_{\mathrm{CPP}}}T_{\mathrm{sym}}$. Let $x_{n,m}$ be the transmitted symbol carried on the $m$-th subcarrier within the $n$-th AFDM symbol. We define $\mathbf{x}_n=\left[ x_{n,0},x_{n,1},...,x_{n,M-1} \right] ^{\mathrm{T}}$ and the corresponding modulated waveform vector as $\mathbf{s}_n=\left[ s_{n,0},s_{n,1},...,s_{n,M-1} \right] ^{\mathrm{T}}$. As derived in \cite{AFDM}, the modulated signal vector for the $n$-th AFDM symbol can be expressed as
\begin{equation}
    \mathbf{s}_n=\mathbf{\Lambda }_{c1}^{\mathrm{H}}\mathbf{F}_{M}^{\mathrm{H}}\mathbf{\Lambda }_{c2}^{\mathrm{H}}\mathbf{x}_n,
\end{equation}
where $\mathbf{\Lambda }_c=\operatorname{diag}\left( e^{-j2\pi cm^2},m=0,1,...,M-1 \right) $, and $\mathbf{F}_M$ denotes the $M\times M$ DFT matrix.
The received signal $r(t)$ through the double-selective channel at the MS is given by
\begin{equation}
r\left( t \right) =\iint{}h\left( \tau ,\nu \right) e^{j2\pi \nu \left( t-\tau \right)}s\left( t-\tau \right) \mathrm{d}\tau \mathrm{d}\nu +n\left( t \right),
\end{equation}
where the DD channel response is defined as 
\begin{equation}
h\left( \tau ,\nu \right) =\sum_{p=1}^P{\alpha _p\boldsymbol{\delta }\left( \tau -\tau _p \right) \boldsymbol{\delta }\left( \nu -\nu _p \right) },
\end{equation}
where $\tau_p$ and $\nu_p$ denote the delay, and Doppler frequency of $p$th propagation path, respectively. 
$n\left( t \right)$ denotes the additive white Gaussian noise, each element follows an i.i.d. complex Gaussian distribution $\mathcal{C} \mathcal{N} \left( 0,\sigma _{\mathrm{n}}^{2} \right) $.
The normalized delay and Doppler are defined as
\begin{equation}
 d_p\triangleq\frac{\tau_p}{T_{\rm s}}=\ell_p+\iota_p,
 \qquad
 \kappa_p\triangleq\nu_pT,
 \label{eq:path_parameters}
\end{equation}
where $\ell_p\in\{0,\ldots,M_{\rm CPP}\}$ and $\iota_p\in[-1/2,1/2)$. In particular, $\ell_p$ is obtained by rounding $d_p$, so that $\iota_p=d_p-\ell_p$. After CPP removal and DAFT demodulation, the received signal at $n$th AFDM symbol is given by
\begin{equation}
 \mathbf y_n=\mathbf H_n\mathbf x_n+\mathbf v_n,
 \qquad
 \mathbf v_n\sim\mathcal{CN}(\mathbf0,\sigma _{\mathrm{n}}^{2}\mathbf I_M),
 \label{eq:afdm_input_output}
\end{equation}
where the effective channel matrix $\mathbf H_n$ is given by
\begin{equation}
 \mathbf H_n=
 \sum_{p=1}^{P}\alpha_p e^{j2\pi\nu_p nT_{\rm sym}}
 \mathbf B(\ell_p,\iota_p,\kappa_p),
 \label{eq:time_varying_afdm_channel}
\end{equation}
where $\mathbf B(\ell_p,\iota_p,\kappa_p)$ denotes the AFDM channel response of $p$th propagation path.

\subsection{AFDM Input--Output Relationship}

Substituting the time-varying channel in \eqref{eq:time_varying_afdm_channel} into \eqref{eq:afdm_input_output}, the AFDM input--output relation can be expressed as
\begin{equation}
 y_n[m]=\sum_{p=1}^{P}\alpha_p e^{j2\pi\nu_p nT_{\rm sym}}
 \sum_{m'=0}^{M-1}B_{p,m,m'}x_n[m']+v_n[m],
 \label{eq:scalar_afdm_input_output}
\end{equation}
where $B_{p,m,m'}$ denotes the $(m,m')$ entry of $\mathbf B(\ell_p,\iota_p,\kappa_p)$ for the $p$-th propagation path, which is given by
\begin{align}
 B_{p,m,m'}={}&e^{-j2\pi c_2m^2}
 e^{j2\pi c_1d_p^2}
 e^{j2\pi(c_2m'^2-d_pm'/M)}
 \mathcal Q_{p,m,m'},
 \label{eq:exact_afdm_coefficient}
\end{align}
where the amplitude response $\mathcal Q_{p,m,m'}$ for fractional delay and Doppler of the pth propagation path is given by \cite{bemani2023afdm}
\begin{align}
 \mathcal Q_{p,m,m'}={}&\frac{1}{M}
 \sum_{q=0}^{C}e^{j2\pi q\iota_p}
 \sum_{\substack{[r-d_p]_M\in\mathcal L_{m,q}}}
 \nonumber\\[-1mm]
 &\quad \times e^{j\frac{2\pi}{M}
 \left(m'-(m+\ell_p^{\rm eq})\right)r},
 \label{eq:exact_qsum}
\end{align}
where $C\triangleq 2Mc_1$ and $\mathcal L_{m,q}$ denote the number of phase-transition regions and the associated index sets induced by the AFDM chirp periodicity, respectively.
The equivalent delay $\ell_p^{\rm eq}$ of $p$th propagation path is defined as $\ell_p^{\rm eq}\triangleq
 [\kappa_p+2Mc_1\ell_p]_M$.

\subsection{Pilot-Aided Observation Model}

Let $B=CM_{\rm CPP}$ denotes the guard interval between the pilot and the data. The first AFDM symbol contains two pilots at $p_1=B$ and $p_2=M-1$. For each pilot, the receiver retains
\begin{equation}
 \mathcal O_j=\{p_j-B,\ldots,p_j\},\qquad j\in\{1,2\},
 \label{eq:dual_pilot_windows}
\end{equation}
and stacks the two windows into $\mathbf y_0^{\rm obs}\in\mathbb C^{2W}$, where $W=B+1$. Let $\mathcal P_0=\{p_1,p_2\}$ and $\mathbf x_{\mathcal P_0}$ represent the known pilot positions and associated pilot values, respectively. The normalized pilot response for each propagation path is given by
\begin{equation}
 \mathbf c_0(\ell,\iota,\kappa)=
 \mathbf B_{\mathcal O_1\cup\mathcal O_2,\mathcal P_0}
 (\ell,\iota,\kappa)\mathbf x_{\mathcal P_0}.
 \label{eq:first_pilot_response}
\end{equation}
To further exploit the information in the time domain and improve the estimation accuracy of Doppler,
one pilot at $M-1$ is also transmitted in the tracking symbols indexed by $\mathcal T=\{n_1,\ldots,n_{N_{\rm tr}}\}$. Let $\mathbf c_{\rm t}(\ell,\iota,\kappa)$ denote its $W$-sample response. The complete known-pilot observation and normalized path response are
\begin{align}
 \mathbf y_J &\triangleq
 \left[(\mathbf y_0^{\rm obs})^T,
 (\mathbf y_{n_1}^{\rm tr})^T,\ldots,
 (\mathbf y_{n_{N_{\rm tr}}}^{\rm tr})^T\right]^T,
 \label{eq:joint_observation}\\
 \mathbf c_J(\ell,\iota,\kappa) &\triangleq
 \begin{bmatrix}
 \mathbf c_0(\ell,\iota,\kappa)\\
 e^{j2\pi(\kappa/T)n_1T_{\rm sym}}\mathbf c_{\rm t}(\ell,\iota,\kappa)\\
 \vdots\\
 e^{j2\pi(\kappa/T)n_{N_{\rm tr}}T_{\rm sym}}\mathbf c_{\rm t}(\ell,\iota,\kappa)
 \end{bmatrix}.
 \label{eq:joint_response}
\end{align}
Thus, the joint observation contains $N_J=(2+N_{\rm tr})W$ samples. Numerical values of the design and stopping parameters are specified only in the simulation setup.

\section{Extrema and Identifiability of the AFDM Atom}
\label{sec:atom_extrema}
\label{subsec:atom_extrema}

In this section, we analyze the extrema of the AFDM response and then examine the identifiability of the delay and Doppler parameters. 

For algorithm development, the discrete normalized response generated by one candidate propagation path over the selected observation rows is referred to as an \emph{AFDM atom}. 
To expose the dependence of $\mathcal Q_{p,m,m'}$ on the mismatch coordinate $\Delta_p$, we decompose its inner summation into the following $q$th geometric component:
\begin{align}
G_{q,p}(\Delta _p)&=
\sum_{[r-(\ell_p+\iota_p)]_M\in\mathcal L_{m,q}}
e^{j\frac{2\pi}{M}\Delta _pr}
\\
&=e^{j\frac{2\pi}{M}\Delta _p\left( a_{q,p}+\frac{K_{q,p}-1}{2} \right)}\frac{\sin \left( \frac{\pi K_{q,p}}{M}\Delta _p \right)}{\sin \left( \frac{\pi}{M}\Delta _p \right)},
 \label{eq:qsum_closed_form}
\end{align}
where $\Delta_p \triangleq m'-(m+\ell_p^{\rm eq})$, while $a_{q,p}$ and $K_{q,p}$ denote the first index and the number of indices, respectively, in the considered contiguous summation segment. 
For an individual contiguous
segment, $\left|G_{q,p}(\Delta_p)\right|
 \le K_{q,p}$, and equality is attained at $\Delta_p=0$. 
Hence, each individual geometric term has a
Dirichlet-type main lobe centered at the coherent point
$\Delta_p=0$, accompanied by sidelobes caused by the finite summation interval.

The complete amplitude response, however, is the phase-weighted combination of all phase-transition terms. Define $\theta_p\triangleq e^{j2\pi\iota_p}$ and $F_p(\Delta_p)\triangleq\sum_{q=0}^{C}\theta_p^qG_{q,p}(\Delta_p)$.
It follows from \eqref{eq:exact_qsum} that
\begin{equation}
 \mathcal Q_{p,m,m'}=\frac{1}{M}F_p(\Delta_p).
\end{equation}
Therefore, the extrema of the AFDM-atom magnitude are determined by $J_p(\Delta_p)
 \triangleq
 \left|F_p(\Delta_p)\right|^2.$
Direct differentiation of the geometric sums with respect to the mismatch coordinate gives
\begin{align}
 F_p'(\Delta_p)
 &=
 j\beta
 \sum_{q=0}^{C}\theta_p^q
 \sum_{[r-(\ell_p+\iota_p)]_M\in\mathcal L_{m,q}}
 r e^{j\beta\Delta_p r},
 \label{eq:qsum_first_delta_derivative}\\
 F_p''(\Delta_p)
 &=
 -\beta^2
 \sum_{q=0}^{C}\theta_p^q
 \sum_{{[r-(\ell_p+\iota_p)]_M\in\mathcal L_{m,q}}}
 r^2 e^{j\beta\Delta_p r},
 \label{eq:qsum_second_delta_derivative}
\end{align}
where $\beta=2\pi/M$. Consequently,
\begin{align}
 &J_p'(\Delta_p)
 =
 2\mathrm{Re}\left\{
 F_p'(\Delta_p)F_p^*(\Delta_p)
 \right\}
 \nonumber\\
 &\quad=
 -2\beta\mathrm{Im}\Bigg\{
 \Bigg[
 \sum_{q=0}^{C}\theta_p^q
 \sum_{[r-(\ell_p+\iota_p)]_M\in\mathcal L_{m,q}}
 r e^{j\beta\Delta_p r}
 \Bigg]
 F_p^*(\Delta_p)
 \Bigg\}.
 \label{eq:atom_first_delta_derivative}
\end{align}
A stationary point $\Delta_p^\star$ therefore satisfies
\begin{align}
 \mathrm{Im}\Bigg\{
 \Bigg[
 \sum_{q=0}^{C}\theta_p^q
 \sum_{[r-(\ell_p+\iota_p)]_M\in\mathcal L_{m,q}}
 r e^{j\beta\Delta_p^\star r}
 \Bigg]
 F_p^*(\Delta_p^\star)
 \Bigg\}
 =0.
 \label{eq:atom_stationary_condition}
\end{align}
The type of stationary point can be determined from
\begin{equation}
 J_p''(\Delta_p)
 =
 2\left|F_p'(\Delta_p)\right|^2
 +
 2\mathrm{Re}\left\{
 F_p''(\Delta_p)F_p^*(\Delta_p)
 \right\}.
 \label{eq:atom_second_delta_derivative}
\end{equation}
A stationary point is a local maximum when
$J_p''(\Delta_p^\star)<0$. At a boundary where the membership of
the summation in $G_{q,p}(\Delta_p)$ changes, the response is treated piecewise
and the extremum is determined from the corresponding one-sided
values.

The individual geometric terms are coherent at $\Delta_p=0$, but
this point is not necessarily a stationary point of their weighted
sum. Since $G_{q,p}(0)=K_{q,p}$, evaluating
\eqref{eq:atom_first_delta_derivative} at $\Delta_p=0$ gives

\begin{align}
 J_p'(0)
 =
 -\frac{4\pi}{M}
 \mathrm{Im}\Bigg\{
 &\Bigg[
 \sum_{q=0}^{C}\theta_p^q
 \sum_{[r-(\ell_p+\iota_p)]_M\in\mathcal L_{m,q}}r
 \Bigg]
 \nonumber\\
 &{}\times
 \left[
 \sum_{q'=0}^{C}\theta_p^{-q'}K_{q',p}
 \right]
 \Bigg\}.
 \label{eq:atom_derivative_at_zero}
\end{align}
When $\iota_p=0$, all the weights $\theta_p^q$ are aligned.
The two bracketed quantities in
\eqref{eq:atom_derivative_at_zero} are then real, and
$J_p'(0)=0$. Thus, the complete response is concentrated around
the nominal coherent location. For a nonzero fractional delay,
the factors $\theta_p^q$ rotate the phase-transition components
relative to one another. Since the cardinalities and first moments of the corresponding index sets generally differ, $J_p'(0)$ is not necessarily zero. The maximum of the complete
response can therefore be displaced from $\Delta_p=0$, even
though every individual geometric term is maximized there.

For pilot $p_j$, let $\Delta_p^\star$ denote a dominant
continuous maximum of $J_p(\Delta_p)$. The associated receive
location approximately satisfies 

\begin{equation}
 m_{p,j}^{(0)}
 \simeq
 \left[
 p_j-\ell_p^{\rm eq}-\Delta_p^\star
 \right]_M.
 \label{eq:dominant_peak_location}
\end{equation}
Since the receive index is sampled only at integer positions, the
observed maximum appears at one of the integer samples surrounding
this continuous location. The dominant-peak coordinate therefore
provides a coarse indication of the integer-delay location. It
does not, by itself, provide an exact estimate of the fractional
delay because $\Delta_p^\star$ also depends on the phase-weighted
interaction among the $q$th-region geometric terms.
\begin{figure}[t]
 \centering
 \includegraphics[width=\linewidth]
 {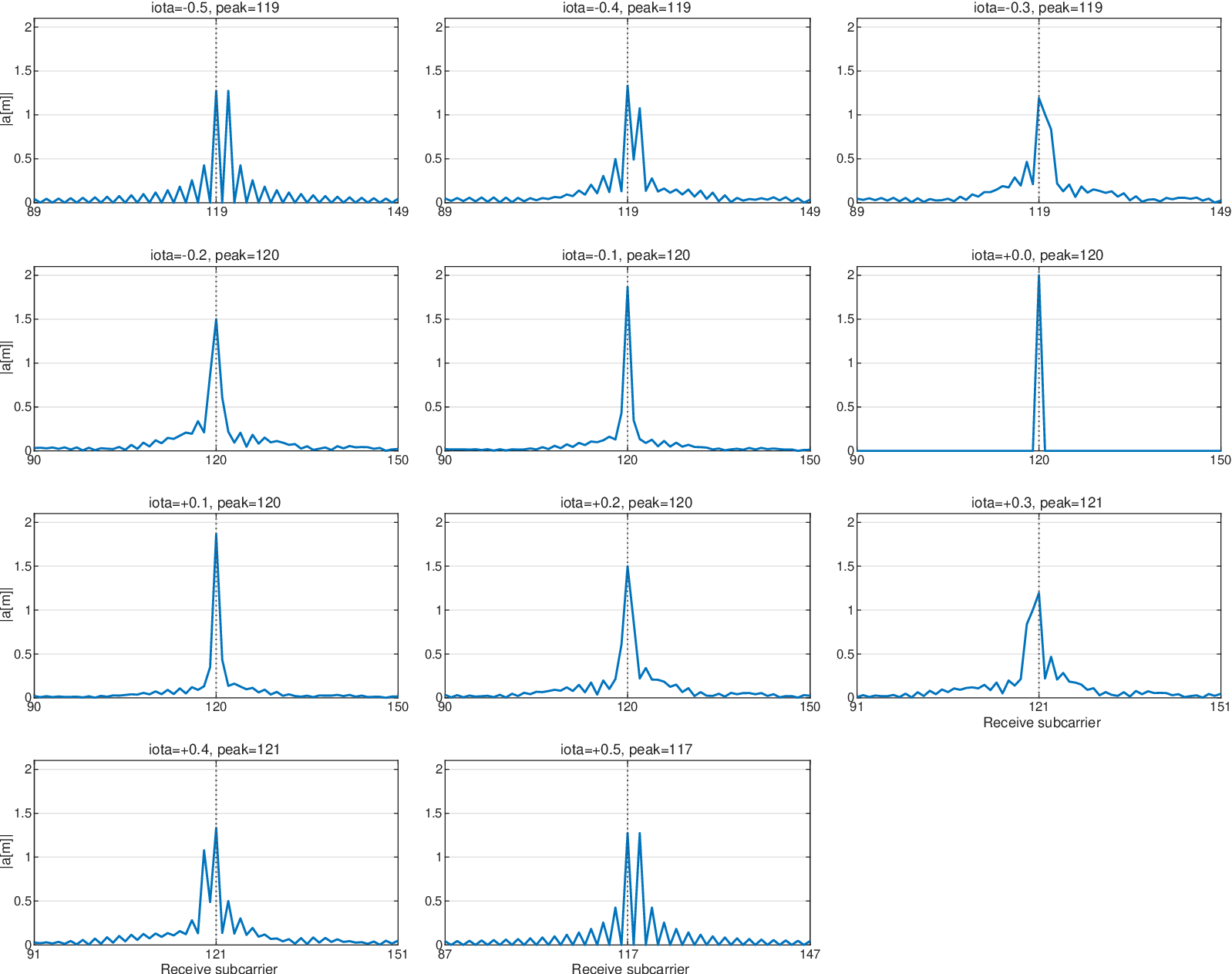}
 \caption{Variation of the sampled magnitude profile of a
 representative AFDM atom with fractional delay. The integer delay,
 Doppler, pilot position, and waveform parameters are fixed. The
 vertical dotted line in each panel marks the strongest sampled
 coefficient.}
 \label{fig:fractional_delay_atom}
\end{figure}

Fig.~\ref{fig:fractional_delay_atom} illustrates the dependence of
the atom profile on $\iota_p$. When $|\iota_p|$ is small, the
phase-transition components remain nearly aligned, and most of the
atom energy is concentrated around the nominal receive location.
As $|\iota_p|$ increases, the relative phases
$\theta_p^q$ rotate, causing partial cancellation around
$\Delta_p=0$ and constructive combination at neighboring
mismatch coordinates. The resulting magnitude profile becomes
asymmetric and its energy spreads over a wider range of receive
indices. Secondary extrema become more pronounced, and two
comparable sampled peaks may appear as $\iota_p$ approaches the
boundary of its admissible interval. This behavior explains why
the dominant peak is suitable for coarse integer-delay
localization, whereas the dominant and secondary complex
coefficients together carry the information required for
off-grid fractional-delay refinement.

\section{Proposed Channel Estimation Algorithm}
\label{sec:proposed_ce}

In this section, we first formulate the channel estimation problem with unknown model order. Then, we develop a model-order-adaptive SAGE framework \cite{Fessler1994SAGE} tailored to AFDM channel estimation. Based on the residual observation, a new path is detected and its delay, Doppler, and complex gain are estimated, followed by sequential refinement of the existing paths. Next, we introduce the adaptive model-order selection criterion. Finally, we examine the monotonicity and computational complexity of the proposed algorithm.

\subsection{Unknown-Order Channel Estimation and the Proposed SAGE Framework}
\label{subsec:profile_objective}
In this subsection, we introduce the proposed SAGE channel estimation framework.
Based on the pilot structure, we model the received signal of AFDM and formulate the channel estimation problem.
For a candidate model order $P$, the received signal can be represented by
\begin{equation}
 \mathbf y=\mathbf C_P(\boldsymbol\Xi_P)\boldsymbol\alpha_P+\mathbf v,
 \
 \boldsymbol\Xi_P=\{\boldsymbol\eta_p\}_{p=1}^{P},
 \
 \boldsymbol\eta_p=[\ell_p,\iota_p,\kappa_p]^T,
 \label{eq:parametric_observation}
\end{equation}
where $\mathbf y$, $\mathbf C_P(\boldsymbol\Xi_P)$, and $\boldsymbol\alpha_P$ denote the pilot observation used in the current estimation stage, the matrix of exact AFDM atoms, and the vector of complex path gains, respectively.
Under circularly symmetric complex Gaussian noise, the negative log-likelihood, up to an additive constant, is
\begin{equation}
 \mathcal L_P(\boldsymbol\Xi_P,\boldsymbol\alpha_P)
 =\frac{\|\mathbf y-\mathbf C_P(\boldsymbol\Xi_P)
 \boldsymbol\alpha_P\|_2^2}{\sigma _{\mathrm n}^{2}}.
 \label{eq:negative_log_likelihood}
\end{equation}
For a residual containing noise only, the normalized correlation at the $k$th search cell is
\begin{equation}
 Z_k=\frac{|\mathbf c_k^H\mathbf v|^2}
 {\sigma_{\mathrm n}^2\|\mathbf c_k\|_2^2}
 \sim {\rm Exp}(1).
 \label{eq:cfar_single_cell_statistic}
\end{equation}
Under the independent-cell approximation, the family-wise false-alarm probability over $K_{\rm eff}$ effective search cells satisfies
\begin{equation}
 P_{\rm FA}=1-\left(1-e^{-\gamma_{\rm ord}}\right)^{K_{\rm eff}},
 \label{eq:cfar_familywise_probability}
\end{equation}
which gives the CFAR threshold \cite{Kay1998Detection,Mamandipoor2016NOMP}
\begin{equation}
 \gamma_{\rm ord}=-\ln\!\left[1-
 \left(1-P_{\rm FA}\right)^{1/K_{\rm eff}}\right],
 \label{eq:cfar_order_threshold}
\end{equation}
where $K_{\rm eff}$ denotes the effective number of approximately independent cells. Hence, \eqref{eq:cfar_familywise_probability} provides an approximate false-alarm calibration when neighboring AFDM atoms are correlated.

Accordingly, the joint model-order and parameter estimation problem is formulated as
\begin{equation}
 \left(\widehat P,\widehat{\boldsymbol\Xi},
 \widehat{\boldsymbol\alpha}\right)
 =\mathop{\arg\min}_{\substack{0\le P\le P_{\max},\\
 \boldsymbol\Xi_P,\boldsymbol\alpha_P}}
 \mathcal F(P,\boldsymbol\Xi_P,\boldsymbol\alpha_P),
 \label{eq:cfar_penalized_problem}
\end{equation}
where
\begin{equation}
 \mathcal F(P,\boldsymbol\Xi_P,\boldsymbol\alpha_P)
 =\mathcal L_P(\boldsymbol\Xi_P,\boldsymbol\alpha_P)
 +\gamma_{\rm ord}P.
 \label{eq:cfar_penalized_objective}
\end{equation}
The first term measures the normalized data mismatch, whereas the second term represents the CFAR evidence required by each retained path. Since $P=\|\boldsymbol\alpha_P\|_0$, \eqref{eq:cfar_penalized_objective} can also be viewed as an $\ell_0$-penalized likelihood. The upper bound $P_{\max}$ specifies the largest admissible model order.

By exploiting the dominant-extremum structure and temporal phase variations of the AFDM pilot responses, we develop an adaptive channel estimation framework based on SAGE tailored to the proposed pilot structure.
Based on the previously estimated channel parameters, we first reconstruct the contributions of the accepted paths.
According to \eqref{eq:parametric_observation}, we have
\begin{equation}
 \mathbf r^{(i)}=\mathbf y-\mathbf C(\widehat{\boldsymbol\Xi}^{(i)})
 \widehat{\boldsymbol\alpha}^{(i)},
 \label{eq:sage_total_residual}
\end{equation}
where $i$ and $\mathbf r^{(i)}$ denote the iteration index and the residual of the current multipath model, respectively.
Next, we use the residual $\mathbf r^{(i)}$ to detect an unresolved propagation path and then estimate its delay, Doppler, and complex gain.
The candidate path can be incorporated into the current model by exploiting the statistically controlled detection criterion.

To resolve the interference between propagation paths, we exploit the reconstructed path contributions to perform a sequential refinement of the channel parameters.
For an accepted path $p$, the path-specific observation in the expectation step (E-step) can be represented by
\begin{equation}
 \mathbf z_p^{(i)}=\mathbf y-
 \sum_{j\ne p}\widehat\alpha_j^{(i)}
 \mathbf c(\widehat{\boldsymbol\eta}_j^{(i)}),
 \label{eq:sage_estep_observation}
\end{equation}
where $\mathbf z_p^{(i)}$ represents the received observation with the reconstructed contributions of all other paths removed.
The channel parameters are estimated from the path-specific observation in the maximization step (M-step).
Given the exact AFDM atom $\mathbf c(\boldsymbol\eta)$, the complex gain can be estimated via
\begin{equation}
 \widehat\alpha(\boldsymbol\eta)=
 \frac{\mathbf c^H(\boldsymbol\eta)\mathbf r}
 {\mathbf c^H(\boldsymbol\eta)\mathbf c(\boldsymbol\eta)+\epsilon},
 \label{eq:profile_gain}
\end{equation}
where $\mathbf r$ and $\epsilon$ denote the single-path observation and the numerical regularization parameter, respectively.
Then, according to \eqref{eq:profile_gain}, we have
\begin{align}
 J(\boldsymbol\eta)
 &=\|\mathbf r-\widehat\alpha(\boldsymbol\eta)
 \mathbf c(\boldsymbol\eta)\|_2^2 \nonumber\\
 &=\|\mathbf r\|_2^2-
 \frac{|\mathbf c^H(\boldsymbol\eta)\mathbf r|^2}
 {\|\mathbf c(\boldsymbol\eta)\|_2^2+\epsilon}.
 \label{eq:profile_cost}
\end{align}
From \eqref{eq:profile_gain} and \eqref{eq:profile_cost}, it can be seen that the single-path parameter estimation problem can be solved by maximizing the normalized matched energy.
Finally, the complex gains of all accepted paths can be obtained as
\begin{equation}
 \widehat{\boldsymbol\alpha}=
 (\mathbf C^H\mathbf C+\lambda\mathbf I)^{-1}\mathbf C^H\mathbf y,
 \label{eq:joint_gain_ls}
\end{equation}
where $\lambda\ge0$ is a numerical diagonal loading and approaches zero in the least-squares gain refits.

In the following subsection, we focus on the estimation of delays, Doppler frequencies, and channel gains.
Then, by exploiting the statistical properties of the residual and the conditional path contributions, we develop the model-order selection criteria for the proposed SAGE framework.

\subsection{Single-Path Delay, Doppler, and Gain Estimation}
\label{subsec:mstep_single_path}

In this subsection, we focus on the estimation of delays, Doppler frequencies, and channel gains from the path-specific observations. The peak location provides a coarse bin estimate. Temporal phase variations provide a coarse Doppler estimate. These estimates determine the candidate integer delays. Finally, the continuous parameters can be obtained by maximizing the profile likelihood, where the complex gain is estimated via \eqref{eq:profile_gain} and jointly refined via \eqref{eq:joint_gain_ls}.

\subsubsection{Dominant-Extremum Delay Initialization}
\label{subsec:peak_search}

Let $\mathbf r_1$ and $\mathbf r_2$ be the two $W$-sample residuals corresponding to the dual pilots. For pilot $p_j$, let $b=p_j-m$ denote the bin coordinate relative to the pilot. Using zero-based window indices, the combined dual-pilot energy at candidate bin $b$ is defined as
\begin{equation}
S(b)=|r_1(W-1-b)|^2+|r_2(W-1-b)|^2,
\qquad 0\le b\le B.
\label{eq:location_energy}
\end{equation}
Let $b_0$ denote the candidate bin that maximizes $S(b)$.
For an interior peak, three-point parabolic interpolation \cite{Werner2016QuadraticInterpolation} gives
\begin{equation}
 \widetilde\rho=b_0+
 \frac{S(b_0-1)-S(b_0+1)}
 {2[S(b_0-1)-2S(b_0)+S(b_0+1)]},
 \label{eq:parabolic_location}
\end{equation}
where the interpolation correction is restricted to $[-1/2,1/2]$. Equation~\eqref{eq:parabolic_location} provides a coarse bin estimate $\widetilde\rho$. The q-sum interference can shift this peak from the true effective-delay location. Together with the coarse Doppler estimate, $\widetilde\rho$ determines the candidate integer delays below. To avoid repeated detection, we exclude neighborhoods of accepted and pruned locations with half-widths $K_{\rm acc}$ and $K_{\rm tabu}$, respectively, as well as previously tested peaks.

\subsubsection{Temporal Doppler Initialization and Integer-Delay Separation}
\label{subsec:coarse_doppler}

We then focus on the estimation of Doppler frequencies. Let $m_{\rm pk}=\operatorname{round}(M-1-\widetilde\rho)$ denote the receive index corresponding to the detected peak for the pilot at $M-1$. Based on the received samples at this index, the pilot-normalized temporal response can be represented by
\begin{equation}
 \mathbf g(\widetilde\rho)=
 [g_0,g_{n_1},\ldots,g_{n_{N_{\rm tr}}}]^T,
 \label{eq:temporal_peak_sequence}
\end{equation}
where $\{0\}\cup\mathcal T$ denotes the symbol offsets associated with the available pilot observations. Given the single-tone nature of the temporal response, the Doppler frequency can be estimated via the maximum-likelihood criterion
\begin{equation}
 \widetilde\nu=\arg\max_{|\nu|\le\nu_{\max}}
 \left|\sum_{n\in\{0\}\cup\mathcal T}
 g_ne^{-j2\pi\nu nT_{\rm sym}}\right|^2.
 \label{eq:coarse_doppler}
\end{equation}
In this paper, a common $N_\nu$-point temporal steering matrix is applied to estimate the Doppler frequencies of all paths. Then the continuous Doppler estimate can be recovered by performing local polynomial interpolation around the largest periodogram sample.

Using the coarse bin and Doppler estimates, we obtain the integer-delay center as
\begin{equation}
 \ell_{\rm c}=
 \operatorname{clip}_{[0,M_{\rm CPP}]}
 \left\{\operatorname{round}
 \left(\frac{\widetilde\rho-\widetilde\nu T}{2Mc_1}\right)\right\}.
 \label{eq:integer_delay_initialization}
\end{equation}
We include the two neighboring integer delays to form the candidate set:
\begin{equation}
 \mathcal L=\{\ell_{\rm c}-1,\ell_{\rm c},\ell_{\rm c}+1\}
 \cap[0,M_{\rm CPP}].
 \label{eq:integer_delay_candidates}
\end{equation}
The candidate set is expanded to neighboring integer delays when the fractional-delay estimate approaches the interval boundary $\pm1/2$ or the normalized residual remains large.

\subsubsection{Joint Fractional-Delay and Doppler Refinement}
\label{subsec:single_path_refinement}

We then estimate the fractional delay and coarse Doppler from the polynomial structure of the AFDM response.
Let $\theta=e^{j2\pi\iota}$. After cancellation of the other paths and the other pilot response, strong peak samples are assumed to be dominated by the path under update. Using the current gain and delay estimates, we normalize these samples as
\begin{align}
 z_{j,m}={}&
 \frac{r_{j,m}-\widehat\alpha
 \widehat B_{m,p_{\bar j}}x_{p_{\bar j}}}
 {\widehat\alpha x_{p_j}}\,
 M e^{j2\pi c_2m^2}e^{-j2\pi c_1\widehat d^2}
 \nonumber\\
 &{}\times e^{-j2\pi c_2p_j^2}e^{j2\pi\widehat d p_j/M},
 \label{eq:normalized_theta_observation}
\end{align}
where $\bar j=3-j$ indexes the other pilot, $\widehat d=\widehat\ell+\widehat\iota$ is the current total-delay estimate, and $\widehat B$ is evaluated at the current path parameters.
Neglecting noise and cancellation errors, \eqref{eq:normalized_theta_observation} gives the local approximation
\begin{equation}
 z_{j,m}\approx\sum_{q=0}^{C}G_{q,j,m}\theta^q,
 \label{eq:theta_measurement}
\end{equation}
where $G_{q,j,m}$ is the inner geometric sum in \eqref{eq:exact_qsum} for $m'=p_j$, evaluated at the current path parameters and held fixed during root finding.
To estimate the coarse fractional delay, we solve
\begin{equation}
 \sum_{q=0}^{C}G_{q,j,m}\theta^q-z_{j,m}=0.
 \label{eq:theta_polynomial}
\end{equation}
By solving the polynomial in \eqref{eq:theta_polynomial} and selecting the admissible root closest to the unit circle, we obtain the coarse estimate of the fractional delay $\iota$.

For each pilot, the admissible sample-wise roots are combined into a coarse generator estimate, denoted by $\bar\theta_j$.
Based on the recovered $\bar\theta_1$ and $\bar\theta_2$, the corresponding delay correction can be estimated as
\begin{equation}
 \Delta d=
 \frac{M\angle(\bar\theta_1\bar\theta_2^*)}
 {2\pi(p_1-p_2)}.
 \label{eq:two_pilot_delay_correction}
\end{equation}
For an admissible correction $|\Delta d|<\epsilon_d$, the corrected coarse fractional-delay estimates are
\begin{equation}
 \iota_j=
 \operatorname{wrap}_{[-1/2,1/2)}
 \left(\frac{\angle\bar\theta_j}{2\pi}
 +\frac{\Delta d\,p_j}{M}\right).
 \label{eq:pilotwise_iota}
\end{equation}

We refine $u\in\{\iota,\kappa\}$ using the full-window profile cost in \eqref{eq:profile_cost}. With the other path parameters fixed, its first and second derivatives are
\begin{align}
 g_u&=\frac{\partial J(\boldsymbol\eta)}{\partial u},\label{eq:newton_gradient}\\
 H_u&=\frac{\partial^2 J(\boldsymbol\eta)}{\partial u^2}.
 \label{eq:newton_curvature}
\end{align}
For $H_u>0$, the Newton direction is $s_u=-g_u/H_u$; otherwise, a bounded negative-gradient direction is used. The step is limited by $\Delta_{u,\max}$ and reduced as $s_u/\beta_{\rm ls}^{b}$, $b=0,\ldots,B_{\rm ls}$, until the full-window cost decreases. Refinement stops if no trial is accepted or $|s_u|<\epsilon_u$.

Following the $N_\iota$-point initialization, we perform $I_{\theta\kappa}$ alternating updates, followed by at most $I_\iota$ and $I_\kappa$ scalar Newton steps. The lowest-cost state and its atom are retained.

Without temporal observations, an $N_\iota^{\rm GN}\times N_\kappa^{\rm GN}$ grid initializes the alternating updates, followed by at most $I_{\rm GN}$ Gauss--Newton iterations. With the gain fixed at its current estimate, Gauss--Newton linearizes the full-window residual through the atom derivatives \cite[Sec.~10.3]{Nocedal2006NumericalOptimization}:
\begin{equation}
 \mathbf e=\mathbf r-\widehat\alpha\mathbf c,\qquad
 \mathbf D=
 \left[\widehat\alpha\frac{\partial\mathbf c}{\partial\iota},
 \widehat\alpha\frac{\partial\mathbf c}{\partial\kappa}\right],
 \label{eq:gn_jacobian}
\end{equation}
Their real-valued forms are
\begin{equation}
 \widetilde{\mathbf D}=
 \begin{bmatrix}\Re\{\mathbf D\}\\\Im\{\mathbf D\}\end{bmatrix},
 \qquad
 \widetilde{\mathbf e}=
 \begin{bmatrix}\Re\{\mathbf e\}\\\Im\{\mathbf e\}\end{bmatrix}.
 \label{eq:real_gn_system}
\end{equation}
The joint update is
\begin{equation}
 \Delta\boldsymbol\eta_c=
 (\widetilde{\mathbf D}^T\widetilde{\mathbf D}+\mu\mathbf I)^{-1}
 \widetilde{\mathbf D}^T\widetilde{\mathbf e},
 \quad
 \boldsymbol\eta_c=[\iota,\kappa]^T,
 \label{eq:gn_step}
\end{equation}
where $\mu=\eta_\mu\max\{\operatorname{diag}(\widetilde{\mathbf D}^T\widetilde{\mathbf D}),\epsilon\}$ is the regularization level. Both components are bounded and tested by backtracking.

For both Newton and Gauss--Newton updates, a trial is accepted only if it decreases
\begin{equation}
 J_{\rm accept}=
 \|\mathbf r-\widehat\alpha_{\rm trial}
 \mathbf c_{\rm trial}\|_2^2.
 \label{eq:full_window_acceptance}
\end{equation}
with $\widehat\alpha_{\rm trial}$ re-estimated over the full window. The derivatives are evaluated by central differences in the implementation.

The single-path parameter estimator is summarized in Algorithm~\ref{alg:single_path_estimation}.
\begin{algorithm}[t]
\caption{Single-Path Delay, Doppler, and Gain Estimation}
\label{alg:single_path_estimation}
\begin{algorithmic}[1]
\REQUIRE Path-specific dual-pilot residuals $\mathbf r_1$ and $\mathbf r_2$, temporal observations, and system parameters
\STATE Construct the dual-pilot location score and obtain $\widetilde\rho$ from \eqref{eq:parabolic_location}.
\STATE Form $\mathbf g(\widetilde\rho)$ using \eqref{eq:temporal_peak_sequence} and estimate $\widetilde\nu$ from \eqref{eq:coarse_doppler}.
\STATE Determine $\ell_{\rm c}$ using \eqref{eq:integer_delay_initialization} and construct $\mathcal L$ according to \eqref{eq:integer_delay_candidates}.
\FOR{each $\ell\in\mathcal L$}
    \STATE Form the normalized pilot samples using \eqref{eq:normalized_theta_observation}.
    \STATE Solve \eqref{eq:theta_polynomial} and retain the admissible roots closest to the unit circle.
    \STATE Obtain the coarse fractional delay using \eqref{eq:two_pilot_delay_correction} and \eqref{eq:pilotwise_iota}.
    \STATE Estimate the gain from \eqref{eq:profile_gain} and evaluate the profile cost in \eqref{eq:profile_cost}.
    \STATE Refine $(\iota,\kappa)$ using \eqref{eq:newton_gradient} and \eqref{eq:newton_curvature}, subject to \eqref{eq:full_window_acceptance}.
\ENDFOR
\STATE Retain the admissible candidate with the minimum profile cost.
\ENSURE $\widehat\ell$, $\widehat\iota$, $\widehat\kappa$, $\widehat\alpha$, and the corresponding exact AFDM atom
\end{algorithmic}
\end{algorithm}

\subsubsection{Fixed-Delay Temporal Doppler and Gain Refinement}
\label{subsec:tracking_refinement}

In this subsection, we focus on the final Doppler and gain refinement with $\widehat P$, $\widehat\rho_p$, and $\widehat d_p$ determined by Section~\ref{subsec:stage1_births}. Let $\mathbf G_{\rm t}\in\mathbb C^{W\times\widehat P}$ denote the matrix of exact unit-gain single-pilot responses. Given the fixed path geometry, the tracking observations can be represented by a common response matrix with time-varying gains. Based on the estimated effective delay and integer delay, the geometric Doppler initializer can be recovered as
\begin{equation}
 \nu_{p,\rm geo}=
 \frac{1}{T}\operatorname{wrap}_{[-1/2,1/2)}
 \{\widehat\rho_p-2Mc_1\widehat\ell_p\}.
 \label{eq:geometric_doppler}
\end{equation}
According to the temporal path-gain model, we have
\begin{equation}
 a_{p,n}^{\rm pred}=
 \widehat\alpha_{p,0}e^{j2\pi\nu_{p,\rm geo}nT_{\rm sym}}.
 \label{eq:predicted_tracking_gain}
\end{equation}

To resolve the uncertainty in tracking gains, we exploit a weakly regularized preliminary LS estimate to perform an innovation and effective-noise assessment. Let $\mathbf Q=\mathbf G_{\rm t}^H\mathbf G_{\rm t}$ denote the common tracking Gram matrix. According to the tracking observation model, we have
\begin{align}
 \widetilde{\mathbf a}_n
 &=(\mathbf Q+\epsilon_g\mathbf I)^{-1}
 \mathbf G_{\rm t}^H\mathbf y_n^{\rm tr},\label{eq:weak_tracking_ls}\\
 \widehat\sigma_{\rm eff,n}^2
 &=\max\left\{\frac{\|\mathbf y_n^{\rm tr}
 -\mathbf G_{\rm t}\widetilde{\mathbf a}_n\|_2^2}{W},
 \sigma _{\mathrm{n}}^{2}\right\}.\label{eq:effective_tracking_noise}
\end{align}
where $\epsilon_g>0$ is a small Tikhonov regularization parameter used to stabilize the preliminary LS estimate \cite{Neumaier1998Regularization}.
The innovation variance $q_p$ is estimated from the median squared difference between $\widetilde a_{p,n}$ and $a_{p,n}^{\rm pred}$. Specifically, the preliminary LS noise bias is subtracted and a positive floor is applied to the resulting variance. Based on the previously estimated innovation variance and effective noise power, the adaptive regularizer can be estimated as
\begin{equation}
 \lambda_{p,n}=
 \operatorname{clip}\left\{
 \frac{\widehat\sigma_{\rm eff,n}^2}{q_p},
 \lambda_{\min},\lambda_{\max}\right\}.
 \label{eq:adaptive_tracking_lambda}
\end{equation}
With the adaptive regularizer, the tracking gains can be estimated as
\begin{equation}
 \widehat{\mathbf a}_n=
 (\mathbf Q+\boldsymbol\Lambda_n)^{-1}
 \left(\mathbf G_{\rm t}^H\mathbf y_n^{\rm tr}
 +\boldsymbol\Lambda_n\mathbf a_n^{\rm pred}\right),
 \label{eq:tracking_gain_ls}
\end{equation}
where $\boldsymbol\Lambda_n=\operatorname{diag}(\lambda_{1,n},\ldots,\lambda_{\widehat P,n})$ denotes the path-dependent regularization matrix.
With the estimated delays fixed, we further refine each path Doppler using \eqref{eq:coarse_doppler} with $g_n$ replaced by the gain-SNR-weighted tracking gains, retaining $\nu_{p,\rm geo}$ when the temporal estimate is unreliable.

Finally, the time-varying channel matrix can be obtained as
\begin{equation}
 \widehat{\mathbf H}_n=
 \sum_{p=1}^{\widehat P}\widehat\alpha_{p,0}
 e^{j2\pi\widehat\nu_pnT_{\rm sym}}
 \mathbf B(\widehat\ell_p,\widehat\iota_p,\widehat\kappa_p).
 \label{eq:estimated_frame_channel}
\end{equation}

\subsection{Adaptive Model-Order Selection}
\label{subsec:stage1_births}

In this subsection, we determine the unknown number of propagation paths through sequential detection and pruning. The single-path estimator in Section~\ref{subsec:single_path_refinement} supplies the candidate parameters, while the CFAR test and path-strength threshold control the number of retained components.

\subsubsection{Sequential Residual-Domain Path Births}

Let $\mathbf m=\mathbf C\widehat{\boldsymbol\alpha}$ denote the reconstructed observation. Starting with $P=0$ and $\mathbf m=\mathbf0$, we search the first-symbol dual-pilot residual $\mathbf r=\mathbf y_0^{\rm obs}-\mathbf m$ and estimate each candidate by the methods in the preceding subsection. Its effective location $\widehat\rho=[\widehat\kappa+2Mc_1\widehat\ell]_M$ must be separated from the accepted and previously pruned locations by at least $\Delta_{\rm acc}$ and $\Delta_{\rm tabu}$, respectively.

The normalized peak statistic is compared with the CFAR threshold in \eqref{eq:cfar_order_threshold}. Equivalently, a candidate is accepted when its contribution produces the normalized residual reduction
\begin{equation}
 \Delta_{\rm b}=\mathcal L_P-\mathcal L_{P+1}
 \ge \gamma_{\rm ord}.
 \label{eq:cfar_birth_acceptance}
\end{equation}
The atom matrix is then augmented, and all gains are jointly updated by \eqref{eq:joint_gain_ls}.

The search terminates at $P_{\max}$ or when either residual test holds:
\begin{align}
 \|\mathbf r\|_2^2&\le
 \epsilon_{\rm rel}\|\mathbf y_0^{\rm obs}\|_2^2,
 \label{eq:relative_residual_stop}\\
 \frac{\|\mathbf r\|_2^2}{\sigma _{\mathrm{n}}^{2}}
 &\le F_\Gamma^{-1}(1-P_{\rm FA,res};2W),
 \label{eq:residual_noise_stop}
\end{align}
where $F_\Gamma^{-1}(\cdot;2W)$ denotes the quantile function of a Gamma distribution with shape parameter $2W$ and unit scale, corresponding to the normalized energy of $2W$ independent circularly symmetric complex Gaussian noise samples \cite{Kay1998Detection}.
Here, $\epsilon_{\rm rel}>0$ is the relative-residual tolerance, and the second test is applied only after at least $P_{\min}$ paths have been accepted.

Subsequent pruning uses the normalized path strength
\begin{equation}
 G_p=\frac{|\widehat\alpha_p|^2\|\mathbf c_p\|_2^2}{\sigma _{\mathrm{n}}^{2}}
 \label{eq:stage1_path_strength}
\end{equation}
with the adaptive threshold
\begin{equation}
 z_{\rm th}=\min\left\{\gamma_{\rm ord},z_{\max},
 \max\left\{z_{\min},\eta_{\rm b}G_{\max}^{\rm birth}\right\}\right\},
 \label{eq:adaptive_stage1_threshold}
\end{equation}
where $G_{\max}^{\rm birth}$ is the largest path strength observed during detection, $\eta_{\rm b}$ sets its relative contribution, and $z_{\min}$ and $z_{\max}$ bound the threshold. The CFAR term ensures $z_{\rm th}\le\gamma_{\rm ord}$.

\subsubsection{SAGE Refinement and Support Pruning}
\label{subsec:sage_round}

The tentative model is refined over at most $I_{\max}$ SAGE rounds. For each path $p$, the contributions of the other paths are subtracted to obtain
\begin{equation}
 \mathbf r_p=\mathbf y_0^{\rm obs}-
 \left(\mathbf m-\widehat\alpha_p\mathbf c_p\right).
 \label{eq:sage_hidden_data}
\end{equation}
The single-path estimator is initialized from the current parameters and applied to $\mathbf r_p$ in the prescribed refinement rounds $\mathcal I_{\rm ref}$. A trial is accepted only if it decreases the profile residual, moves the effective location by no more than $\Delta_{\rm move}$, and maintains a distance greater than $\Delta_{\rm sep}$ from the other paths. These distances use $d_M(a,b)=|[a-b+M/2]_M-M/2|$. The reconstructed observation is updated as
\begin{equation}
 \mathbf m\leftarrow\mathbf m-
 \widehat\alpha_p^{\rm old}\mathbf c_p^{\rm old}
 +\widehat\alpha_p^{\rm new}\mathbf c_p^{\rm new}.
 \label{eq:incremental_model_update}
\end{equation}
All gains are jointly re-estimated after each path sweep.

Paths satisfying $G_p<z_{\rm th}$ are removed, subject to the minimum order $P_{\min}$, and their locations are recorded to prevent immediate rediscovery. The gains are re-estimated after pruning. If the path set changes, the next round starts from the reduced model; otherwise, convergence is declared when
\begin{equation}
 \frac{\|\widehat{\boldsymbol\alpha}^{(i)}
 -\widehat{\boldsymbol\alpha}^{(i-1)}\|_2}
 {\|\widehat{\boldsymbol\alpha}^{(i)}\|_2+\epsilon}
 <\epsilon_{\rm SAGE},
 \label{eq:sage_convergence}
\end{equation}
where $\epsilon>0$ prevents division by zero and $\epsilon_{\rm SAGE}$ is the convergence tolerance.

Finally, all known pilots are jointly used to recover missed paths, refine the retained parameters, and prune components with insufficient conditional contributions before the fixed-support Doppler and gain refinement in Section~\ref{subsec:tracking_refinement}.

The adaptive model-order SAGE procedure is summarized in Algorithm~\ref{alg:stage1_sage}.
\begin{algorithm}[t]
\caption{Proposed SAGE Estimation With Adaptive Model Order}
\label{alg:stage1_sage}
\begin{algorithmic}[1]
\REQUIRE Dual-pilot observation, tracking pilots, $\sigma_{\mathrm n}^{2}$, and system parameters
\STATE Form the stacked observation $\mathbf y$, set $P=0$, and compute $\gamma_{\rm ord}$ using \eqref{eq:cfar_order_threshold}.
\WHILE{$P<P_{\max}$ and neither residual stopping test holds}
    \STATE Generate a separated candidate from the dual-pilot residual using Algorithm~\ref{alg:single_path_estimation}.
    \STATE Jointly refit the gains and accept the birth only if \eqref{eq:cfar_birth_acceptance} holds.
\ENDWHILE
\STATE Compute $G_p$ using \eqref{eq:stage1_path_strength} and set $z_{\rm th}$ using \eqref{eq:adaptive_stage1_threshold}.
\FOR{$i=1,\ldots,I_{\max}$}
    \FOR{$p=1,\ldots,P$ and $i\in\mathcal I_{\rm ref}$}
        \STATE Form \eqref{eq:sage_hidden_data}, refine path $p$, and retain only a cost-decreasing separated trial.
    \ENDFOR
    \STATE Jointly refit all gains, prune paths satisfying $G_p<z_{\rm th}$, and refit the retained gains.
    \STATE Stop if the support is unchanged and \eqref{eq:sage_convergence} holds.
\ENDFOR
\STATE Use all known pilots for missed-path recovery, joint refinement, and conditional pruning.
\STATE Apply the fixed-support Doppler and gain refinement in Section~\ref{subsec:tracking_refinement}.
\ENSURE $\widehat P$ and $\{\widehat\ell_p,\widehat\iota_p,\widehat\nu_p,\widehat\alpha_{p,0}\}_{p=1}^{\widehat P}$
\end{algorithmic}
\end{algorithm}

\section{Monotonicity of the Penalized Objective}
\label{sec:monotonicity}

The following result establishes the monotonicity of the objective in \eqref{eq:cfar_penalized_objective} during the CFAR-controlled SAGE iterations.

\begin{proposition}[Monotonicity of the penalized objective]
\label{prop:penalized_descent}
Suppose that $\gamma_{\rm ord}>0$ is fixed during one estimation run and that the gain refitting is performed by least squares. Under the birth rule in \eqref{eq:cfar_birth_acceptance}, the residual stopping rules in \eqref{eq:relative_residual_stop} and \eqref{eq:residual_noise_stop}, the cost-decreasing SAGE update, and the pruning rule $G_p<z_{\rm th}\le\gamma_{\rm ord}$, the objective sequence generated by the proposed estimator satisfies
\begin{equation}
 0\le \mathcal F^{(i+1)}\le \mathcal F^{(i)}.
 \label{eq:penalized_monotone_chain}
\end{equation}
\end{proposition}

The proof is provided in Appendix~\ref{app:penalized_descent_proof}.

\section{Computational Complexity}
\label{subsec:proposed_complexity}

In this section, the computational complexity of the proposed algorithm is analyzed. The main complexity lies in repeated exact-atom evaluation, observation-domain matrix products for gain estimation, and temporal Doppler estimation. Let $N_J=(2+N_{\rm tr})W$ denote the maximum observation dimension and $P$ the largest model order encountered during estimation. With a fixed number of pilots per symbol, one exact-atom evaluation takes at most $\mathcal O(CN_J)$ operations. The Gram-matrix construction $\mathbf C^H\mathbf C$ in \eqref{eq:joint_gain_ls} takes $\mathcal O(N_JP^2)$ operations, while the corresponding $P$-dimensional solve is lower order when $N_J\gg P$. The temporal steering-matrix multiplication in \eqref{eq:coarse_doppler} takes $\mathcal O(PN_\nu(1+N_{\rm tr}))$ operations for all paths, where $N_\nu$ is the number of Doppler grid points. Finally, the leading-order complexity is bounded by
\begin{equation}
 \mathcal O\!\left(
 N_J\bigl(CK_{\rm a}+K_{\rm g}P^2\bigr)
 +K_{\rm t}PN_\nu(1+N_{\rm tr})
 \right),
 \label{eq:total_estimation_complexity}
\end{equation}
where $K_{\rm a}$, $K_{\rm g}$, and $K_{\rm t}$ count the atom evaluations, observation-domain gain fits, and temporal Doppler updates, respectively. Scalar operations and low-dimensional parameter updates are omitted.
Table~\ref{tab:estimator_complexity_comparison} gives the total online orders for the implemented AFDM estimators.

\begin{table*}[!t]
\caption{Computational Complexity of the AFDM Channel Estimators}
\label{tab:estimator_complexity_comparison}
\centering
\footnotesize
\renewcommand{\arraystretch}{1.25}
\setlength{\tabcolsep}{1.5pt}
\begin{tabular}{@{}>{\centering\arraybackslash}p{0.16\textwidth}|*{4}{>{\centering\arraybackslash}p{0.202\textwidth}|}@{}}
\hline
\textbf{Estimator} & \multicolumn{4}{c}{\textbf{Computational complexity}} \\
\hline
Proposed (this work) & Exact-atom evaluation & Joint gain fitting & Temporal Doppler estimation & Total complexity \\
\cline{2-5}
& $\mathcal O(CN_JK_{\rm a})$ & $\mathcal O(N_JK_{\rm g}P^2)$ & $\mathcal O(K_{\rm t}PN_\nu S)$ & $\mathcal O\!\left(\begin{gathered}N_J(CK_{\rm a}+K_{\rm g}P^2)\\ {}+K_{\rm t}PN_\nu S\end{gathered}\right)$ \\
\hline
Fixed-order SAGE \cite{Fessler1994SAGE} & \multicolumn{4}{c}{Total complexity} \\
\cline{2-5}
& \multicolumn{4}{c}{$\mathcal O\!\left(N(CK_{{\rm a},m}+K_{{\rm g},m}P^2)+N_JP^2+(K_{{\rm t},m}+1)PN_\nu S\right)$} \\
\hline
NOMP-CFAR \cite{Mamandipoor2016NOMP} & \multicolumn{4}{c}{Total complexity} \\
\cline{2-5}
& \multicolumn{4}{c}{$\mathcal O\!\left(N(CK_{{\rm a},m}+K_{{\rm g},m}P^2)+N_JP^2+(K_{{\rm t},m}+1)PN_\nu S\right)$} \\
\hline
SAGE-BIC \cite{Fessler1994SAGE,Xiong2017BIC} & \multicolumn{4}{c}{Total complexity} \\
\cline{2-5}
& \multicolumn{4}{c}{$\mathcal O\!\left(N(CK_{{\rm a},m}+K_{{\rm g},m}P_{\max}^2+P_{\max}^3)+N_JP^2+(K_{{\rm t},m}P_{\max}+P)N_\nu S\right)$} \\
\hline
VAMP \cite{Rangan2019VAMP} & \multicolumn{4}{c}{Total complexity} \\
\cline{2-5}
& \multicolumn{4}{c}{$\mathcal O\!\left(I_{\rm V}(ND+N^2)+N_JP^2+WCP+PN_\nu S\right)$} \\
\hline
MDVALSE-inspired \cite{Zhang2020MDVALSE} & \multicolumn{4}{c}{Total complexity} \\
\cline{2-5}
& \multicolumn{4}{c}{$\mathcal O\!\left(N[PD+I_{\rm M}CPL+(P+I_{\rm M})P^2]+N_JP^2+PN_\nu S\right)$} \\
\hline
SBL \cite{Wipf2004SBL} & \multicolumn{4}{c}{Total complexity} \\
\cline{2-5}
& \multicolumn{4}{c}{$\mathcal O\!\left(ND+I_{\rm S}(NQ^2+Q^3)+I_{\rm off}N(CP+P^2)+N_JP^2+PN_\nu S\right)$} \\
\hline
\end{tabular}
\end{table*}

\section{Simulation Results}
\label{sec:simulation_results}
In this section, simulation results are demonstrated to examine the performance of the proposed channel estimation methods and the comparison with benchmarks.
The main simulation parameters are listed in Table~\ref{tab:simulation_parameters}. The SNR is defined as $\mathrm{SNR}\triangleq \frac{\left\| \mathbf{Y} \right\| _{\mathrm{F}}^{2}}{\left\| \mathbf{N} \right\| _{\mathrm{F}}^{2}}$.
The maximum MS velocity is set to 300 km/h and 600 km/h, resulting in a maximum Doppler frequency of $\nu _{\max}=$ 4167 Hz and 8333 Hz. 
The chirp parameters are set as $c_1=1/M$ and $c_2=0$.

\begin{table}[!t]
\caption{Simulation Parameters}
\label{tab:simulation_parameters}
\centering
\renewcommand{\arraystretch}{1.2}
\setlength{\tabcolsep}{4pt}
\begin{tabular}{|>{\centering\arraybackslash}p{0.67\columnwidth}|>{\centering\arraybackslash}p{0.24\columnwidth}|}
\hline
\textbf{Parameter} & \textbf{Value} \\
\hline
DAFT-domain samples per AFDM symbol, $M$ & 1024 \\
\hline
CPP length (maximum integer delay), $M_{\rm CPP}$ & 80 \\
\hline
DAFT chirp parameters, $(c_1,c_2)$ & $(1/M,\,0)$ \\
\hline
One-sided response search width, $B_{\rm half}$ & 160 \\
\hline
Subcarrier spacing, $\Delta f$ (kHz) & 30 \\
\hline
Effective symbol duration, $T=1/\Delta f$ ($\mu$s) & 33.33 \\
\hline
Inter-symbol interval, $T_{\rm sym}$ ($\mu$s) & 35.94 \\
\hline
Carrier frequency, $f_c$ (GHz) & 15 \\
\hline
Maximum Doppler frequency, $\nu_{\max}$ (Hz) & 4167 / 8333 \\
\hline
Maximum terminal speed, $v_{\max}$ (km/h) & 300 / 600 \\
\hline
AFDM symbols per frame & 14 \\
\hline
Modulation  & 16-QAM \\
\hline
Pilot amplitude / Data amplitude & 2 / 1 \\
\hline
Number of propagation paths & [5, 10] \\
\hline
\end{tabular}
\end{table}

We compare the proposed estimator with the following benchmark schemes. All methods are evaluated on the same channel and noise realizations, followed by the same frame-level tracking and data-detection procedure.
\begin{itemize}
\item \textbf{Fixed-order SAGE} \cite{Fessler1994SAGE}: The true path number is supplied, while path parameters are estimated by sequential residual fitting and cyclic leave-one-out refinement. The path gains are updated by joint least squares.
\item \textbf{NOMP-CFAR} \cite{Mamandipoor2016NOMP}: New paths are detected from the residual and refined over continuous delay and Doppler coordinates. Cyclic refinement and joint least squares update the accepted paths, whereas a CFAR threshold controls path addition.
\item \textbf{SAGE-BIC} \cite{Fessler1994SAGE,Xiong2017BIC}: An overcomplete SAGE estimate is first obtained. Nested path subsets are then refitted by least squares, and the Bayesian information criterion selects the model order.
\item \textbf{VAMP} \cite{Rangan2019VAMP}: A coarse AFDM DD dictionary is combined with a Bernoulli--Gaussian denoiser. The resulting activity probabilities and residual evidence determine the selected grid components.
\item \textbf{MDVALSE-inspired} \cite{Zhang2020MDVALSE}: Sequential evidence tests determine the path support, while a local two-dimensional posterior over fractional delay and Doppler yields refined parameter estimates using exact AFDM responses.
\item \textbf{SBL} \cite{Wipf2004SBL}: Sparse Bayesian learning estimates the coefficient variances of a selected coarse-grid atom pool through iterative evidence updates. The selected paths receive limited off-grid delay and Doppler refinement.
\end{itemize}

\begin{figure}[t]
	\centering
	\begin{minipage}[htbp]{1.0\linewidth}
	\centering
	\includegraphics[width=0.85\linewidth]{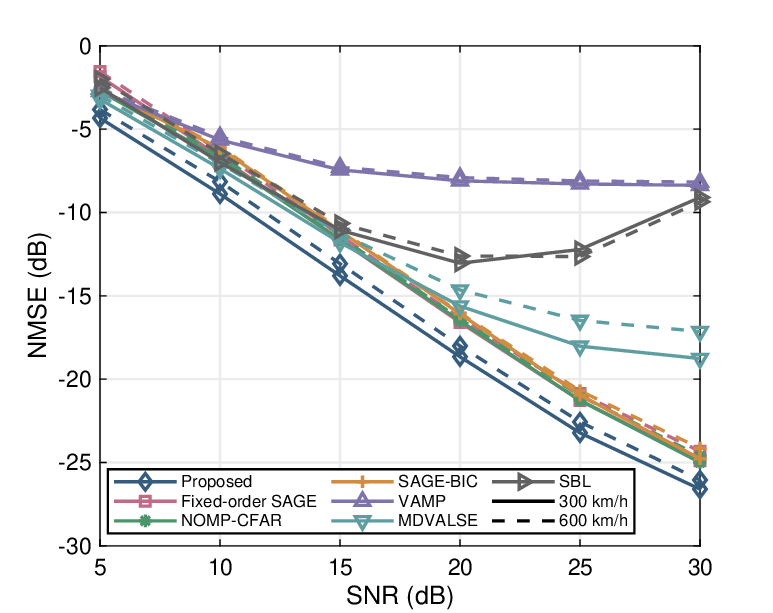}
	\caption{The NMSE performance versus SNR.}
	\label{NMSEvsSNR}
	\end{minipage}%
	\hfill
\end{figure}

The NMSE performance of the proposed method and the benchmark schemes under different SNRs is shown in Fig.~\ref{NMSEvsSNR}. As can be observed, the proposed method consistently achieves the lowest NMSE over the entire SNR range. Fixed-order SAGE, NOMP-CFAR, and SAGE-BIC also exhibit steadily improved performance as the SNR increases, but remain inferior to the proposed method due to less effective parameter initialization or model-order determination. In addition, VAMP, MDVALSE, and SBL exhibit noticeable error floors in the high-SNR regime, mainly because the residual basis mismatch and local refinement errors become dominant as the noise level decreases. Furthermore, when the velocity increases from 300 km/h to 600 km/h, only a slight NMSE degradation is observed for the proposed method, demonstrating its robustness against the more severe Doppler effect in highly time-varying channels. Furthermore, even at 600 km/h, the proposed method still achieves lower NMSE than the benchmark schemes under the same mobility condition, which verifies its effectiveness in extremely high-mobility scenarios.

\begin{figure}[t]
	\centering
	\begin{minipage}[htbp]{1.0\linewidth}
	\centering
	\includegraphics[width=0.85\linewidth]{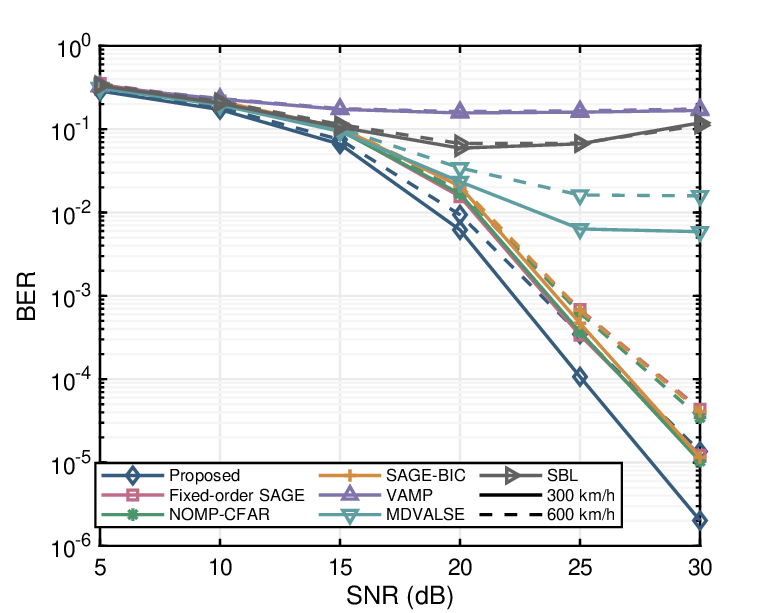}
	\caption{The BER performance versus SNR, the QAM order is 16QAM.}
	\label{BERvsSNR}
	\end{minipage}%
	\hfill
\end{figure}

Fig.~\ref{BERvsSNR} presents the BER performance of the considered methods under 16-QAM modulation. The proposed method achieves the lowest BER over the considered SNR range, which is consistent with the NMSE results in Fig.~\ref{NMSEvsSNR}. 
The superior BER performance mainly benefits from the improved channel estimation accuracy, which reduces the residual channel mismatch during data detection. Fixed-order SAGE, NOMP-CFAR, and SAGE-BIC also exhibit rapidly decreasing BER as the SNR increases, whereas VAMP, MDVALSE, and SBL suffer from evident BER floors due to their relatively large channel estimation errors in the high-SNR regime. Furthermore, increasing the velocity from 300 km/h to 600 km/h leads to only a limited BER degradation for the proposed method. Even under the severe mobility condition of 600 km/h, the proposed method still achieves lower BER than the benchmark schemes evaluated at the same velocity, indicating that its robustness to high Doppler spreads can be effectively translated into reliable data detection performance.

\begin{figure}[t]
	\centering
	\begin{minipage}[htbp]{1.0\linewidth}
	\centering
	\includegraphics[width=0.85\linewidth]{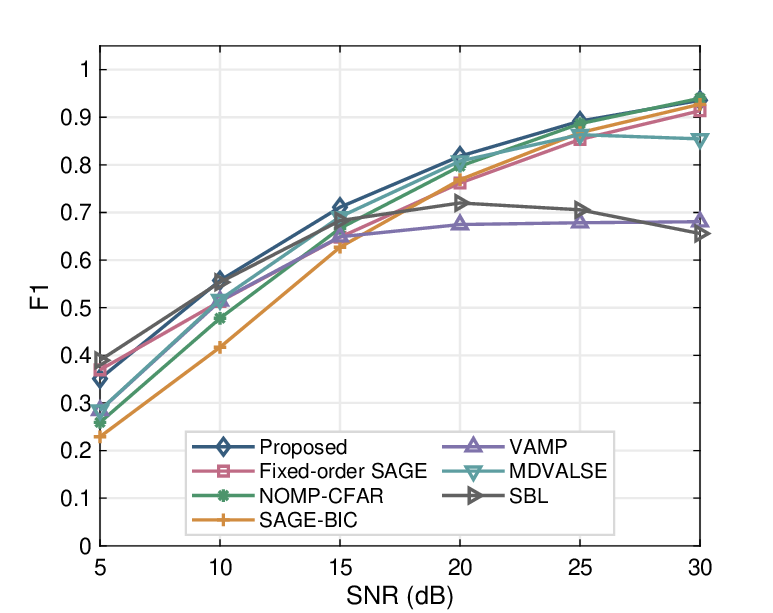}
	\caption{Path support F1 score versus SNR.}
	\label{F1vsSNR}
	\end{minipage}%
	\hfill
\end{figure}

Fig.~\ref{F1vsSNR} illustrates the path support identification performance of the proposed method and the benchmark schemes versus SNR. The F1 score is defined as the harmonic mean of precision and recall, i.e., $F_1=2\mathrm{Prec}\,\mathrm{Rec}/(\mathrm{Prec}+\mathrm{Rec})$, where $\mathrm{Prec}=\mathrm{TP}/(\mathrm{TP}+\mathrm{FP})$ and $\mathrm{Rec}=\mathrm{TP}/(\mathrm{TP}+\mathrm{FN})$. Here, $\mathrm{TP}$, $\mathrm{FP}$, and $\mathrm{FN}$ denote the numbers of correctly identified, falsely detected, and missed paths, respectively. A detected path is counted as correct when its effective-delay location is matched one-to-one with a true path within 1.5 samples. As shown in Fig.~\ref{F1vsSNR}, the proposed method achieves the highest or comparable F1 score in the moderate-to-high-SNR regime, increasing from approximately 0.35 at 5 dB to 0.94 at 30 dB. NOMP-CFAR and SAGE-BIC approach its performance at high SNR, whereas VAMP and SBL exhibit clear F1 floors. The small difference between the proposed method and NOMP-CFAR at 30 dB indicates that both methods recover most path locations under favorable noise conditions.

\begin{figure}[t]
	\centering
	\begin{minipage}[htbp]{1.0\linewidth}
	\centering
	\includegraphics[width=0.85\linewidth]{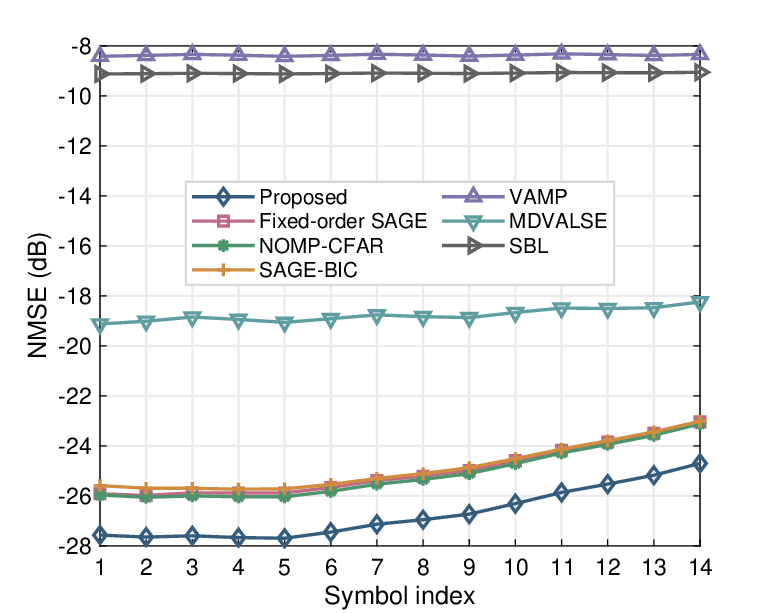}
	\caption{The NMSE performance versus the symbol index. The SNR is 30 dB.}
	\label{NMSEvsSymbol}
	\end{minipage}%
	\hfill
\end{figure}

To evaluate the robustness against temporal channel variation, the NMSE performance over successive symbols is reported in Fig.~\ref{NMSEvsSymbol}. 
The proposed method consistently maintains the best estimation accuracy across all considered symbol indices.
As the symbol index increases, the NMSE of the proposed method gradually increases due to the accumulated phase error over time, which is caused by the Doppler estimation error. 
Nevertheless, the performance degradation remains limited, demonstrating the robustness of the proposed method to channel time variation. Fixed-order SAGE, NOMP-CFAR, and SAGE-BIC exhibit similar trends but consistently yield higher NMSE, while VAMP and SBL suffer from pronounced error floors. Although MDVALSE achieves better performance than VAMP and SBL, a noticeable performance gap still exists compared with the proposed method. 
These results indicate that the temporal Doppler initialization and continuous parameter refinement enable the proposed method to maintain more accurate channel reconstruction over successive symbols.

\begin{figure}[t]
	\centering
	\begin{minipage}[htbp]{1.0\linewidth}
	\centering
	\includegraphics[width=0.85\linewidth]{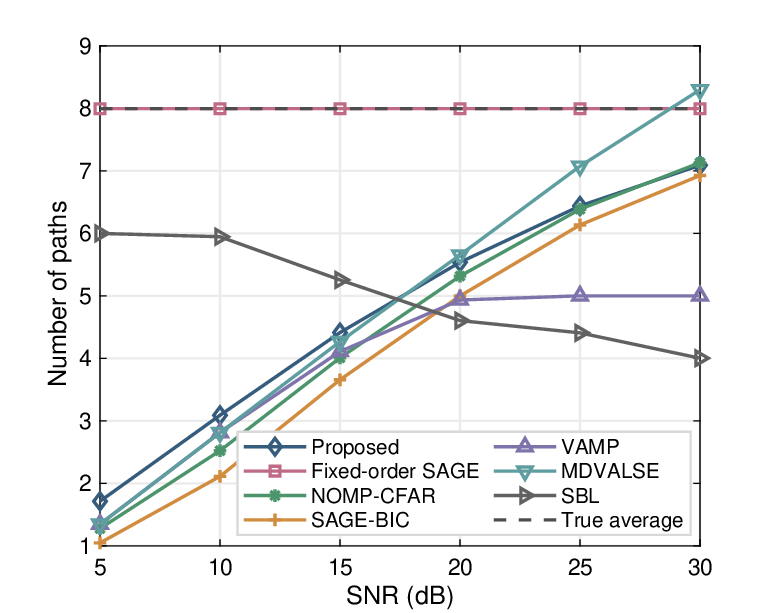}
	\caption{The average number of estimated paths versus SNR.}
	\label{PathNumbervsSNR}
	\end{minipage}%
	\hfill
\end{figure}

Fig.~\ref{PathNumbervsSNR} compares the average number of estimated paths obtained by different methods. 
As the SNR increases, the proposed method detects an increasing number of propagation paths and gradually approaches the true model order, since weak paths become more distinguishable from noise at higher SNRs. 
NOMP-CFAR and SAGE-BIC exhibit a similar tendency, whereas VAMP and SBL tend to underestimate the number of paths even in the high-SNR regime. In contrast, MDVALSE gradually increases the estimated model order and slightly overestimates the true average at high SNR. Fixed-order SAGE remains at eight paths because the true model order is assumed to be known in advance. It is worth noting that the proposed method does not necessarily recover all weak paths even at high SNR, since its adaptive model-order search retains only paths with sufficiently significant contributions. Nevertheless, the resulting channel estimate still achieves the best NMSE and BER performance, as demonstrated in Figs.~\ref{NMSEvsSNR} and \ref{BERvsSNR}, indicating that the proposed model-order selection effectively balances weak-path detection and estimation reliability.

\begin{figure}[t]
	\centering
	\begin{minipage}[htbp]{1.0\linewidth}
	\centering
	\includegraphics[width=0.85\linewidth]{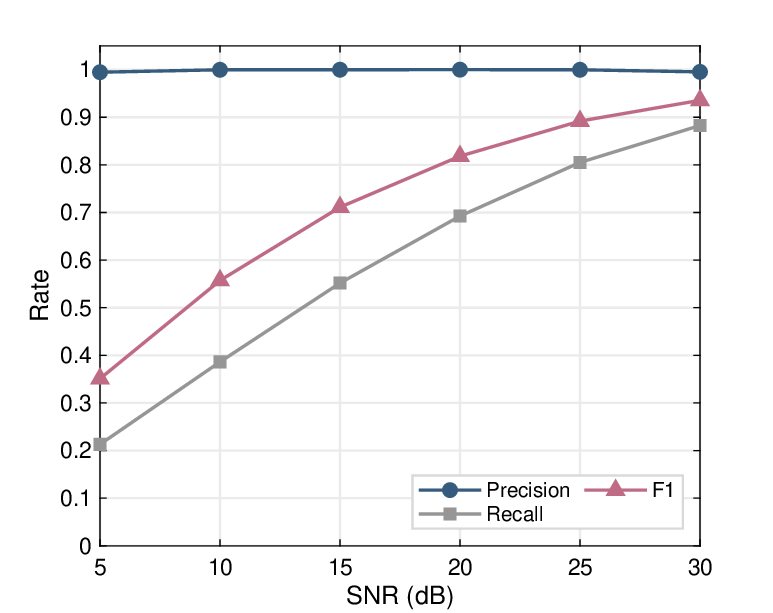}
	\caption{The path detection performance versus SNR.}
\label{PathDetectionvsSNR}
	\end{minipage}%
	\hfill
\end{figure}

Fig.~\ref{PathDetectionvsSNR} illustrates the path detection performance of the proposed method in terms of precision, recall, and F1 score. As shown in Fig.~\ref{PathDetectionvsSNR}, the precision remains close to one over the entire SNR range, indicating that the proposed method introduces very few false paths during the adaptive model-order search. In contrast, the recall gradually increases with SNR, since weak propagation paths become more distinguishable from noise as the observation quality improves. Accordingly, the F1 score also increases steadily and approaches one in the high-SNR regime. These results demonstrate that the proposed method adopts a relatively conservative path-selection strategy at low SNR, while progressively recovering more valid paths as the SNR increases, thereby achieving a favorable balance between false-alarm suppression and weak-path detection.

\begin{figure}[t]
	\centering
	\begin{minipage}[htbp]{1.0\linewidth}
	\centering
	\includegraphics[width=0.85\linewidth]{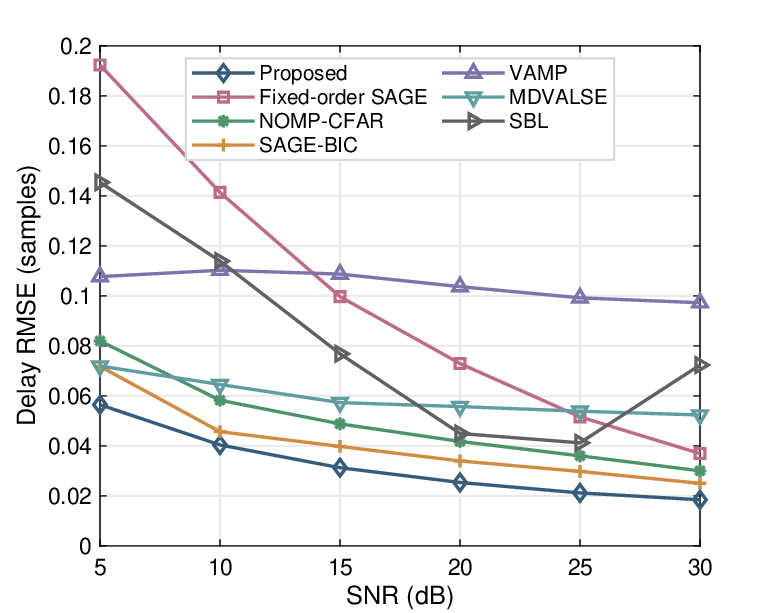}
	\caption{The delay RMSE performance versus SNR.}
	\label{DelayRMSEvsSNR}
	\end{minipage}%
	\hfill
\end{figure}

To further examine the parameter estimation accuracy, Fig.~\ref{DelayRMSEvsSNR} reports the delay RMSE of the proposed method and the benchmark schemes. As shown in Fig.~\ref{DelayRMSEvsSNR}, the proposed method consistently achieves the lowest delay RMSE over the considered SNR range, and its estimation error decreases steadily as the SNR increases. 
This performance improvement mainly benefits from the continuous fractional-delay refinement based on the exact AFDM response, which effectively mitigates the off-grid mismatch associated with discretized delay representations. SAGE-BIC and NOMP-CFAR also exhibit improved delay estimation accuracy with increasing SNR, but remain inferior to the proposed method. In contrast, VAMP and MDVALSE exhibit noticeable error floors in the high-SNR regime, while SBL shows a non-monotonic behavior due to its sensitivity to support estimation and local parameter refinement. These results further verify the effectiveness of the proposed continuous-parameter refinement for fractional-delay estimation.

\begin{figure}[t]
	\centering
	\begin{minipage}[htbp]{1.0\linewidth}
	\centering
	\includegraphics[width=0.85\linewidth]{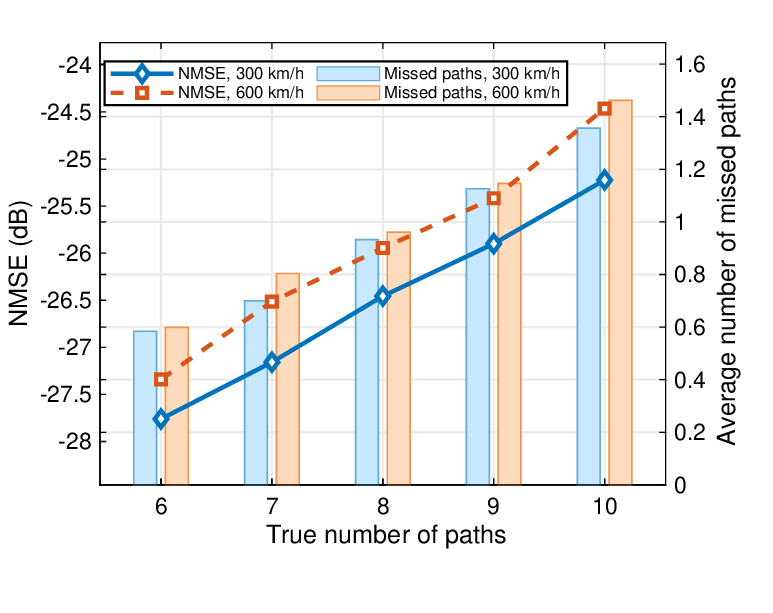}
	\caption{The channel NMSE and average number of missed paths versus the true number of propagation paths. The SNR is 30 dB.}
	\label{PathvsNMSE}
	\end{minipage}%
	\hfill
\end{figure}

\begin{figure}[t]
	\centering
	\begin{minipage}[htbp]{1.0\linewidth}
	\centering
	\includegraphics[width=0.85\linewidth]{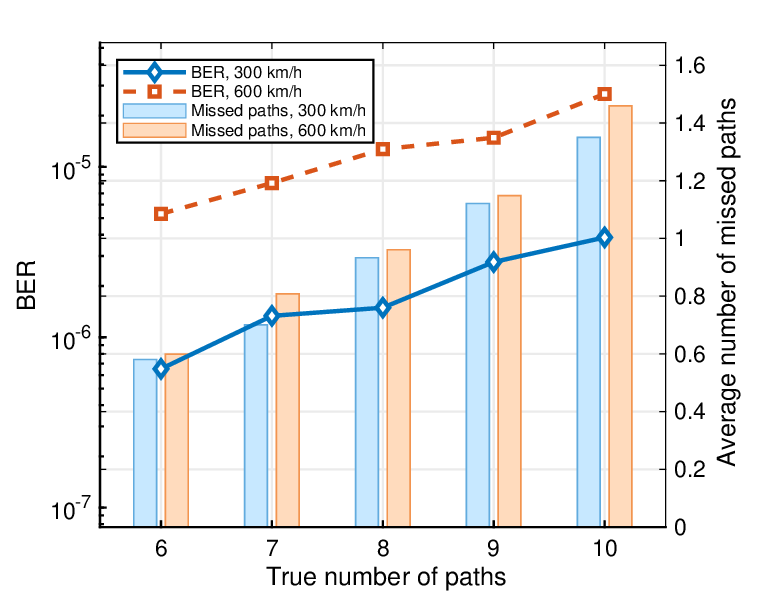}
	\caption{The BER and average number of missed paths versus the true number of propagation paths. The SNR is 30 dB.}
	\label{PathvsBER}
	\end{minipage}%
	\hfill
\end{figure}

Fig.~\ref{PathvsNMSE} shows the relationship between the NMSE and the average number of missed paths at 30 dB for different true path numbers. As the path number increases from 6 to 10, the NMSE performance degrades and more paths are missed at both maximum velocities. Under a fixed pilot budget, the receiver needs to resolve extra DD responses and estimate more path gains from the same received samples. Response overlaps hinder the separation of weak paths, and missing components are not incorporated into the reconstructed channel model. The NMSE at 600 km/h is consistently larger than that at 300 km/h, showing that rapid channel variations raise the estimation complexity.

Fig.~\ref{PathvsBER} illustrates the corresponding BER together with the missed-path statistics. For both mobility scenarios, the BER increases as the true path number increases, and it remains consistently higher at 600 km/h. The bar results are obtained from the same retained frames used for the BER curves, and their upward trend agrees well with the NMSE degradation observed in Fig.~\ref{PathvsNMSE}. Although additional resolvable paths can offer stronger diversity when perfect CSI is available \cite{AFDM}, they simultaneously enlarge the set of unknown channel parameters under a fixed pilot overhead. Consequently, missed paths and inaccuracies in the retained path parameters introduce residual channel mismatch during data detection. The BER trend therefore indicates that, under this practical estimation condition, the degradation in channel accuracy may outweigh the diversity gain promised by a larger number of paths.

\section{Conclusion}
\label{sec:conclusion}

In this paper, we investigated pilot-aided channel estimation for  AFDM systems with fractional delay and Doppler. We characterized the exact off-grid input--output relationship and analyzed the extrema and identifiability of the AFDM path response. Based on these properties, we proposed an adaptive SAGE estimator that combines dual-pilot path initialization, multi-symbol observations, continuous delay--Doppler refinement, and model-order selection. A subsequent temporal refinement reconstructs the time-varying channel over the entire frame. 
Simulation results showed that the proposed method achieves lower NMSE and BER than the considered benchmarks over the tested SNR range and maintains its advantage under high-mobility conditions.

\appendices
\section{Proof of Proposition~\ref{prop:penalized_descent}}
\label{app:penalized_descent_proof}
Assuming that the algorithm has estimated $P$ paths, then we have
\begin{equation}
 \mathcal L_P=
 \frac{\|\mathbf y-\mathbf C_P(\boldsymbol\Xi_P)
 \widehat{\boldsymbol\alpha}_P\|_2^2}{\sigma_{\mathrm n}^2},
 \qquad
 \mathcal F_P=\mathcal L_P+\gamma_{\rm ord}P.
 \label{eq:appendix_penalized_definitions}
\end{equation}
We first consider an accepted path birth. From \eqref{eq:cfar_birth_acceptance}, we have
\begin{align}
 \mathcal F_{P+1}-\mathcal F_P
 &=\mathcal L_{P+1}-\mathcal L_P+\gamma_{\rm ord}\nonumber\\
 &=-\Delta_{\rm b}+\gamma_{\rm ord}
 \le0.
 \label{eq:appendix_birth_descent}
\end{align}
When either condition in \eqref{eq:relative_residual_stop} or \eqref{eq:residual_noise_stop} is satisfied, no new path is introduced. Thus, $\mathcal F_P$ remains unchanged. For a fixed support, the penalty $\gamma_{\rm ord}P$ is constant. Based on \eqref{eq:profile_cost}, a retained SAGE update does not increase the global residual cost when the other path contributions are fixed. Then, the least-squares gain refit cannot increase $\mathcal L_P$.

Next, we examine the effect of path pruning. After the joint least-squares refit, we have $\mathbf C_P=[\mathbf C_{-p},\mathbf c_p]$, where $\mathbf C_{-p}$ contains all atoms except path $p$, and $\mathcal L_{P-1}^{(-p)}$ represents the normalized residual after deleting path $p$ and refitting the retained gains. The orthogonal projector onto the complement of the column space of $\mathbf C_{-p}$ is given by
\begin{equation}
 \mathbf P_{\mathbf C_{-p}}^{\perp}
 =\mathbf I-\mathbf C_{-p}
 (\mathbf C_{-p}^{H}\mathbf C_{-p})^{-1}
 \mathbf C_{-p}^{H}.
 \label{eq:appendix_orthogonal_projector}
\end{equation}
Based on the least-squares deletion identity, the residual variation is given by
\begin{align}
 \Delta_{{\rm d},p}
 &\triangleq \mathcal L_{P-1}^{(-p)}-\mathcal L_P\nonumber\\
 &=\frac{|\widehat\alpha_p|^2}{\sigma_{\mathrm n}^2}
 \mathbf c_p^H\mathbf P_{\mathbf C_{-p}}^{\perp}\mathbf c_p
 \le G_p,
 \label{eq:appendix_deletion_bound}
\end{align}
where the inequality follows from $\mathbf0\preceq\mathbf P_{\mathbf C_{-p}}^{\perp}\preceq\mathbf I$ and the definition of $G_p$ in \eqref{eq:stage1_path_strength}. Since a path is pruned only when $G_p<z_{\rm th}$ and $z_{\rm th}\le\gamma_{\rm ord}$ in \eqref{eq:adaptive_stage1_threshold}, we have
\begin{align}
 \mathcal F_{P-1}^{(-p)}-\mathcal F_P
 &=\Delta_{{\rm d},p}-\gamma_{\rm ord}
 \le G_p-\gamma_{\rm ord}<0.
 \label{eq:appendix_deletion_descent}
\end{align}
Similarly, after each gain refit, consecutive pruning operations do not increase the objective. Thus, every path birth, residual stopping, SAGE refinement, gain refit, and path pruning satisfies
\begin{equation}
 0\le\mathcal F^{(i+1)}\le\mathcal F^{(i)},
 \label{eq:appendix_objective_limit}
\end{equation}
which completes the proof of Proposition~\ref{prop:penalized_descent}.

\bibliographystyle{IEEEtran}
\bibliography{references}

\end{document}